\def\aa{{A\&A}}

\def\aj{{AJ}}

\def\apj{{ApJ}}

\def\mnras{{MNRAS}}

\documentclass[cup5b]{caps}
\usepackage{graphicx}
\usepackage{amssymb}
\usepackage{ociwsymp3}   %Use this for invited papers.

\HeadText{T. Treu}  %Enter author name; if three authors, list all three;
\newcommand{\kms}{km\,s$^{-1}$}
\newcommand{\dv}{$R^{1/4}\,$}

\newcommand{\sbe}{$SB_{\rm e }$}

\newcommand{\Rekpc}{$R_{\rm e }$}

\newcommand{\dlogMLB}{$d \left(\log M_{*}/L_{\rm B} \right) / dz$}

\begin{document}

\pagenumbering{arabic}

%Author names should be in capital letters
\author[]{T. TREU\\California Institute of Technology, Astronomy 105-24, Pasadena, CA, 91115, USA}

\chapter{The Formation of Early-Type Galaxies: Observations to $z\sim1$}

\begin{abstract}
How does the number density of early-type galaxies (E+S0) evolve with
redshift? What are their star formation histories? Do their mass
density profile and other structural properties evolve with redshift?
Answering these questions is key to understanding how E+S0s form and
evolve. I review the observational evidence on these issues, focusing
on the redshift range $z\sim0.1-1$, and compare it to the predictions
of current models of galaxy formation.
\end{abstract}

\section{Introduction}

Understanding the formation and evolution of early-type galaxies
(E+S0, i.e. ellipticals and lenticulars) is not only crucial to unveil
the origin of the Hubble sequence, but is also a focal point
connecting several unanswered major astrophysical questions. The
hypothesis that E+S0s form by mergers of disks at relatively recent
times is one of the pillars of the cold dark matter (CDM) hierarchical
scenario. At galactic scales, since they are the most massive
galaxies, E+S0s are the key to understanding how and when dark and
luminous mass are assembled in galaxies, and to test the universal
form and ubiquity of dark matter halos predicted by the CDM paradigm
(Navarro, Frenk \& White 1997, hereafter NFW; Moore et al.\ 1998). On
subgalactic scales, the existence of a correlation between black-hole
mass and spheroid velocity dispersion suggests that the growth of
black holes and the activity cycles in active galactic nuclei are
somehow intimately connected with the formation of
spheroids. Therefore, a unified formation scenario must ultimately be
conceived (e.~g. Kauffmann \& Haehnelt 2000; Monaco, Salucci \& 
Danese 2000;
Volonteri, Haardt \& Madau 2003; see also the proceedings of meeting I of the
Carnegie Observatories Centennial).

\medskip

Theoretical formation scenarios are often grouped into two categories,
broadly referred to as the {\it monolithic collapse} and {\it
hierarchical formation}\footnote[1]{A complete review of the
theoretical background is beyond the aims of this observational
review. For more information the reader is referred to, e.g., the
reviews by de Zeeuw \& Franx (1991), Bertin \& Stiavelli (1993),
Merritt (1999), Peebles (2002), de Freitas Pacheco, Michard \&
Mohayaee (2003), and references therein.}.

In the traditional picture -- the monolithic collapse -- E+S0s
assembled their mass and formed their stars in a rapid event, of much
shorter duration than their average age (Eggen, Lynden-Bell \& Sandage
1962; Larson 1975; van Albada 1982). The formation process happened at
high redshifts and proto early-type galaxies would be star forming and
dust-enshrouded systems. These kinds of models are consistent with a
variety of features (see Matteucci 2002 and references therein)
including the homogeneity of the present day stellar populations
(Sandage \& Visvanathan 1978), the existence of metallicity gradients
(Sandage 1972) and the characteristic \dv surface brightness profile
(de Vaucouleurs 1948).

By contrast, in the ``hierarchical scenario'' -- hereafter the {\it
standard model} -- early-type galaxies form by mergers of disks at
relatively recent times (Toomre \& Toomre 1972; Toomre 1977; White \&
Rees 1978; Blumenthal et al.\ 1984).  The formation process is
continuous: mass is accreted over time, and both major and minor
mergers can induce star formation thus rejuvenating at times the
stellar populations. Furthermore, environmental processes -- such as
galaxy interactions -- can be built into the models and predictions
made of the properties of E+S0s as a function of environment
(Kauffmann 1996; Benson, Ellis \& Menanteau 2002). Examples that can
be tested against observations include the global properties of E+S0s,
such as the color-magnitude relation or the age of the integrated
stellar populations. Increasingly sophisticated numerical cosmological
simulations are being developed: it has recently become possible to
simulate in detail\footnote{Note however that crucial mechanisms such
as star formation can only be treated in a simplified way by means of
semi-analytical recipes.} the formation of individual E+S0s in a fully
cosmological context (Meza et al.\ 2003).  This opens up the
possibility of using observations of the internal structure (e.g. the
mass density profile) of E+S0s as a test of the {\it standard model}.

A common and practical tool are the so-called pure luminosity
evolution (PLE) models. In these phenomenological models, E+S0s form
at a given redshift of formation ($z_f$) and evolve only through the
evolution of their stellar populations. Typically, the star formation
history is assumed independent of present day luminosity. For a given
star formation history, stellar evolution models, and present day
luminosity function, it is straightforward to compute observable
properties, such as number counts and observed color distribution. PLE
models are often used as toy-realizations of the monolithic collapse
models and their predictions contrasted to the {\it standard model}
predictions. However, it should be kept in mind that PLE models are
only a phenomenological tool, not coincident with monolithic collapse.

\medskip

For decades, the only way to test and improve our understanding of the
formation process was through observations of the local universe. The
only accessible pieces of information were observables such as color
or spectra of local E+S0s.

This has dramatically changed in the last few years. The sharp images
taken by the Hubble Space Telescope (HST), together with the
high-quality ground based data collected by large aperture telescopes
equipped with modern instruments, have opened up the cosmic-time
domain. Now E+S0s can be identified, counted, and studied as a
function of redshift (i.e. cosmic time) out to look-back times that
are a significant fraction of the lifetime of the
Universe. Increasingly detailed information (luminosity, color,
redshift, internal kinematics, mass estimates from dynamics and
lensing) for distant E+S0s can now be obtained, allowing for
increasingly stringent tests of the cosmological model.

Clearly, the combination of both pieces of information -- local and
high redshift data -- is what delivers the most stringent
observational tests. Since other speakers at this meeting have covered
the local Universe (e.g, Davies), I will focus specifically on the
study of distant E+S0s. In particular, I will cover the redshift range
$0.1<z<1$, corresponding in the currently favored $\Lambda$CDM
cosmology\footnote[2]{I assume the Hubble Constant to be
H$_0$=65$h_{65}$~\kms\,Mpc$^{-1}$=100$h$~\kms\,Mpc$^{-1}$, $h_{65}=1$
when needed. The matter density and cosmological constant in critical
units are $\Omega_{\rm m}=0.3$ and $\Omega_{\Lambda}=0.7$,
respectively.} to look-back times of 1 to 8 Gyrs, thus approximately
the second half of the life of the Universe. I will discuss E+S0s in
general, irrespective of their environment. When needed, I will
distinguish between field and cluster E+S0s to contrast different
evolutionary histories. 
A final caveat is
that I will mostly consider the broad class of spheroids (E+S0),
without distinguishing between pure ellipticals and lenticulars. This
simplification should be kept in mind when interpreting observational
results, given that Es and S0s might have significantly different
formation histories (e.g., Dressler et al.\ 1997; Trager 2003). When
possible I will discuss the results in terms of pure Es or S0s.

I will concentrate on three key questions:

\begin{enumerate}
\item How does the number density of E+S0s evolve with time?
\item What is the star formation history of E+S0s?
\item What is the distribution of mass in E+S0s and how does it evolve
with time?
\end{enumerate}

In the next sections, I will review observational work on each of
these questions, discuss comparison with model predictions and briefly
comment on future perspective. 

\section{Evolution of the number density}

How many E+S0s are there at any given redshift? Ideally, in order to
compare directly with models of structure formation, we would like
observations to deliver the volume density of E+S0s as a function of
mass and redshift. The closest available observable to the mass
function is the luminosity function $\phi(L,z)$. If luminosity
evolution is understood, $\phi(L,z)$ can be used to derive the
evolution of the number density at a fixed present day equivalent
luminosity.

Before we proceed, it is useful to introduce a simple parametrization
of $\phi(L,z)$ that can be used to express observational results in a
synthetic form.  Assuming pure luminosity evolution and indicating
stellar mass with $M_{*}$, we can express luminosity evolution as
$\log L(z)=\log L(0) - \left[ d \left(\log M_{*}/L \right) / dz
\right] z $ to first order in $z$. Similarly, assuming that the shape
of the LF is time invariant, the evolution of the overall
normalization $\phi_*$ can be parameterized as
$\phi_*(z)=\phi_*(0)(1+z)^p$.

\subsection{The Luminosity Function of E+S0 galaxies at $z<1$}

\label{sec:LF}

\begin{figure}
    \centering 
\includegraphics[width=7.5cm,angle=0]{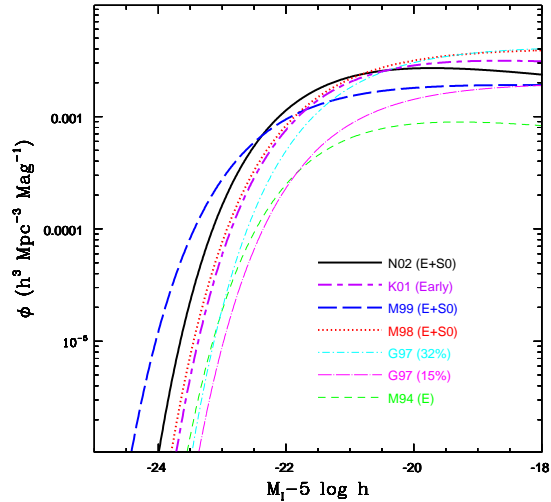}
\caption{Local luminosity functions of E+S0s transformed to
the I-band, see references for details (N03=Nakamura et al.\ 2003,
K01=Kochaneck et al.\ 2001; M99=Marinoni et al.\ 1999; M98=Marzke et
al.\ 1998). Other local LF adopted to construct PLE models are shown
for comparison as thin lines (G97=Gardner et al.\ 1997 total K-band
luminosity function scaled by 0.32 and 0.15; M94=Marzke et al.\ 1994
pure ellipticals luminosity function).}
\label{fig:LFs}
\end{figure}

Before considering distant galaxies, let us briefly summarize our
knowledge of the local LF of E+S0s. To this aim, a compilation is
shown in Figure~\ref{fig:LFs} (heavy lines; see caption for details
and references). The compilation includes LF in various photometric
bands, from the blue to the near infrared. For ease of comparison, the
best fit Schechter (1976) LFs have been transformed to an intermediate
wavelength, using the average colors of E+S0s\footnote[3]{B-I=2.1,
r'-I=0.24 Fugukita, Shimasaku \& Ickikawa (1995); B$_{\rm Z}$-I=2.4 
Im et al.\ (1996);
I-K=2.1 from Bower, Lucey \& Ellis (1992) and Fugukita et al.\ (1995)} to
obtain L$_*$ in the I band. Note that a simple shift in color is only
an approximate transformation, because of the existence of the color
magnitude relation and of different definitions of magnitudes,
photometric system, and morphological classes adopted by various
authors (see, e.g., Kochanek, Keeton \& McLeod 2001).  The agreement among the
most recent determinations is rather encouraging. Nevertheless, as we
will see, the uncertainty in the local LF (the fossil evidence)
contributes significantly to the error budget in the measurement of
$\phi(L,z)$.

The most extensive study of the evolution of the luminosity function
of morphologically selected early-type galaxies is based on the 145
E+S0s with $16.5<I< 22$ in the Groth Strip Survey (Im et al.\ 2002,
hereafter I02; other samples are given in Franceschini et al.\ 1998;
Schade et al.\ 1999; Benson et al.\ 2002; given the morphological
definition adopted here I will not review similar studies based on
spectral classification, e.g. Cohen 2002; Willis et al.\ 2002). The
authors use HST photometry, together with spectroscopic (45/145) and
photometric redshifts (100/145), to compute the rest-frame B band
luminosity, and then proceed to measure the evolution of the LF.

Based on their sample of 145 objects in the redshift range 0.1-1.2
(median redshift 0.6), I02 find \dlogMLB $= -0.76 \pm 0.32$
(i.e. $1.89\pm0.81$ mags of brightening to $z=1$) and $p=-0.86\pm0.68$
(i.e. the number density at $z=1$ is $0.55^{+0.33}_{-0.21}$ the local
value). The large uncertainties arise mostly from the limited size of
the sample, but also from the fact that an apparent magnitude limited
sample probes different volumes and absolute magnitude ranges in the
local and distant universe. Therefore bright E+S0s, dominant at large
z, will have very few counter parts in the local universe and,
viceversa, the faint galaxies dominating the counts in the local
universe will go undetected at large z. I02 try to remove this source
of uncertainty, by fixing the characteristic luminosity of the local
luminosity function to some external measurement based on a larger
volume. Unfortunately, I02 find very different results according to
the local LF they adopt. The Marzke et al.\ (1998) LF yields \dlogMLB
$= -0.79 \pm 0.09$ $p=-0.95\pm0.48$ while the Marinoni et al.\ (1999)
LF yields \dlogMLB $= -0.41 \pm 0.09$ $p=0.12\pm0.54$ (the two LF are
shown in Figure~\ref{fig:LFs} labeled as M98 and M99 respectively).
In conclusion, systematic uncertainties in the local LF hinder
substantial progress.

>From these results it is clear that a two-pronged strategy must be
followed in order to improve on the current factor of 3 uncertainty in
the number density evolution to $z\sim1$. On the one hand, it is
necessary to increase the size of the high redshift sample possibly to
several thousands objects with a larger fraction of spectroscopic
redshifts . Also a sample collected along multiple independent line of
sight will be desirable to minimize the effects of cosmic variance and
clustering of E+S0s (see discussion in I02 and in 1.2.2). The future
prospects appear promising, due to the recent improvements of
observational capabilities both on the imaging side with the Advanced
Camera for Survey on HST, and on the spectroscopic side with the new
generation of wide field high multiplexing spectrographs. On the other
hand, it is necessary to reduce the uncertainty on the local LF of
E+S0s. As illustrated in Figure~\ref{fig:EROs} prospects look good and
hopefully the optical luminosity function per morphological type will
be know with higher accuracy once the morphological classification of
large numbers of galaxies in the Sloan Digitized Sky Survey (SDSS) and
2dF is completed.

\subsection{Extremely Red Objects and the luminosity function of E+S0s at $z\sim1$}

Old stellar populations are characterized by a sharp break in their
spectral energy distribution around 4000 \AA, with larger flux at
longer wavelengths. Hence, an old stellar population at $z\sim1$
appears as an object with extremely red optical to infrared colors,
i.e. and Extremely Red Object (ERO). Therefore, a census of EROs would
provide directly the number density of {\it old} E+S0s at $z\gtrsim 1$
without the need for spectroscopic redshifts, provided contaminants
such as cold stars or dust enshrouded galaxies could be removed.

\begin{figure}
    \centering 
\includegraphics[width=7.5cm,angle=0]{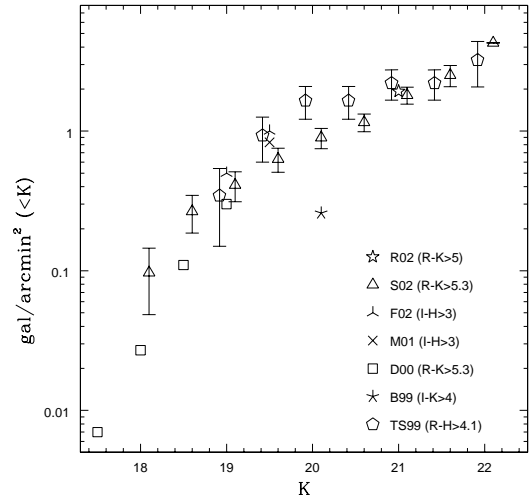}
\caption{Cumulative number density of Extremely Red Objects for
similar color definitions (see references for details; R02=Roche et
al.\ 2002; S02=Smith et al.\ 2002; F02=Firth et al.\ 2002;
M01=McCarthy et al.\ 2001; D00=Daddi et al.\ 2000a; B99=Barger et al.\
1999; TS99=Treu \& Stiavelli 1999). The outlier B99 is measured from
the Hubble Deep Field alone -- which is thought to be deficient in
high redshift E+S0s. This is a clear example of the effects of cosmic
variance and an illustration of the need for a wide field of view and
multiple lines of sight.}
\label{fig:EROs}
\end{figure}

Several optical-infrared surveys over the past five years have been
conducted with the goal of measuring the number density evolution of
E+S0s (Zepf 1997; Moustakas et al.\ 1997; Barger et al.\ 1999; Benitez
et al.\ 1999; Menanteau et al.\ 1999; Treu \& Stiavelli 1999; Thompson
et al.\ 1999; Daddi et al.\ 2000a,b; McCracken et al.\ 2000; Yan et
al. 2000; Corbin et al.\ 2000; Martini 2001; McCarthy et al.\ 2001;
Firth et al.\ 2002; Chen et al.\ 2002; Smith et al.\ 2002; Roche et
al.\ 2002). A selection of the results is shown in
Figure~\ref{fig:EROs}. In spite of the slightly different color cuts
adopted by various groups, it is clear that those surveys conducted
over sufficiently wide areas and/or along multiple lines of sight
(where clustering bias and cosmic variance are unimportant) are in
good agreement.

A comparison with the local abundance of E+S0s can be done considering
simple PLE models. Given a spectral evolution model and the selection
criteria of an ERO survey, we can obtain the set of redshifts
${\mathcal Z(L)}$ at which a galaxy of present day luminosity $L$
would be included in the sample (typically an interval limited on the
low redshift side by the red color criterion and on the high redshift
side by the detection limit). Then, the density of EROs is obtained by
integrating the local luminosity function times the cosmic volume per
unit solid angle $dV/dz$ over the appropriate range in luminosity and
redshift:
\begin{equation}
\int_{L_{\rm min}}^{L_{\rm max}} \int_{\mathcal Z(L)} \phi(L) 
\frac{dV}{dz} dz dL .
\end{equation}
For a local Schechter luminosity function with a flat faint end slope
(c.f. Fig~\ref{fig:LFs}), the model number density depends linearly on
the local characteristic density, while the dependence on the
characteristic luminosity is a rapidly varying function of the depth
of the survey. At $L_*$ -- where the LF is steep -- an uncertainty of
0.3(0.5) mags in the assumed $L_*$ affects the predicted counts at the
50\%(80\%) level. At $L_*/10$ the same uncertainties affect the counts
only at the 10\% (20\%) level. The same calculation can be used to
estimate the effects of the uncertainties in luminosity evolution. It
is clear that in order to perform a reliable comparison with the local
LF it is necessary to go significantly fainter than $L_*$
(corresponding to K=18-18.5 adopting the N02 LF and a reasonable range
of evolutionary models). It is thus necessary to reach beyond
K=20.5-21 to make the uncertainty on $L_*$/luminosity evolution
smaller than the observational errors on the number counts.

The uncertainties related to modeling the star formation history are
more dramatic. It is sufficient to have a small amount of recent star
formation to make the optical to infrared color significantly bluer
and therefore change dramatically ${\mathcal Z(L)}$, and hence the
predicted number counts (see discussion in Jimenez et al.\
1999). Without more information on the star formation history of E+S0s
(see 1.3), it is convenient to adopt the following approach. Models
with no delayed star formation (single burst) will predict the maximum
density of red E+S0s at high-z. A comparison with these models defines
the fraction of local E+S0s already ``old'' at $z\sim1$ (Treu \&
Stiavelli 1999).

\begin{table}
    \begin{tabular}{ccccc} \hline \hline {Obs./Pred.} & {Model LF} & Area (arcmin$^2$) & depth & {Ref}\\
    \hline 
    70 \% & 0.15$\times$K (G97)     & 2200     & H$<21.0$ & McCarthy et al.\ (2001)\\
    100\% & E (M94)                 & 701$^a$  & K$<19.2$ & Daddi et al.\ 
(2000b)\\ 
    33 \% & E+S0 (K01)              & 81.5     & K$<21.0$ & Roche et al.\ (2002)\\ 
    100\% & E (M94)$^b$             & 49$^a$   & K$<21.6$ & Smith et al.\ (2002)\\
    25 \% & 0.32$\times$K (G97)     & 13.8$^a$ & H$<23.2$ & Treu \& Stiavelli (1999)\\
\hline
\hline
\end{tabular} 
  \caption{Number density of EROs with respect to the prediction of
  PLE models. Column Obs./Pred. lists the fraction of predicted
  galaxies that is observed, i.~e. observed/predicted. Model LF lists
  the local LF assumed in the PLE models (Notation as in
  Fig.~\ref{fig:LFs}).(a) Area is a function of depth; I report here
  the maximum area and depth (see references for details). (b) PLE
  models from Daddi et al.\ (2000b).}
\label{tab:EROf}
\end{table}

How do the EROs counts compare with PLE models? I summarize some of
the observational results in Table~\ref{tab:EROf} (the deepest and
widest for which I could find a PLE comparison). At first glance the
results seem highly discrepant. However, some of the fractions in
Table~\ref{tab:EROf} are with respect to the LF of pure ellipticals
(M94 and 0.15K G97; see Fig.~\ref{fig:LFs}) and not of ellipticals
{\it and} lenticulars. If we assume for simplicity that E and S0
galaxies are present approximately in equal numbers at these
luminosities, we have to halve the fractions of McCarthy et al.\
(2001), Daddi et al.\ (2000b) and Smith et al.\ (2002) to obtain the
ratio between the density of EROs and that of local E+S0s (35\%, 50\%,
50\% respectively). The corrected fractions are in much better mutual
agreement and range between 25 and 50\%. This range can be readily
explained in terms of measurement errors, and different star formation
histories and local LF adopted in the PLE models.  A further
correction is required, because we have to take into account the
presence of possible contaminants, such as highly dust enshrouded
starbursts or reddened AGNs (e.g., Smail et al.\ 2002). To address
this, researchers have relied on HST images, to determine what
fraction of EROs are {\it morphologically} E+S0s. A list of the
determinations of morphological fractions among EROs to date is shown
in Table~\ref{tab:EROm}.  It is sufficient to multiply the corrected
fractions from Table~1.1 by the fractions in Table~1.2 to obtain the
density of red E+S0s $z\sim1$. The density of red E+S0s at $z\sim1$ is
8\%-40\% of the local value, with most of the range coming from the
uncertainties in Table~1.2. Even allowing for some extra uncertainty
related to the range of local LF it seems clear that the number of
E+S0s already ``old'' at $z\sim1$ is less than in the local universe.

\begin{table}
    \begin{tabular}{cccc}
     \hline \hline
     {Fraction of E+S0} & {Instrument} & Number of EROs & {Ref}\\
     \hline
     25-35 \% & WFPC2        & 115   & Yan \& Thompson (2003)\\
     20-50 \% & WFPC2        &  60   & Smith et al.\ (2002)\\
     50-80 \% & NICMOS/WFPC2 &  41   & Moriondo et al.\ (2000)\\
     55-75 \% & NICMOS       &  30   & Stiavelli \& Treu (2001)\\
     \hline \hline
    \end{tabular}
\caption{Morphology of EROs. Fraction of EROs with E+S0 morphology
found by various surveys at infrared (NICMOS) or optical (WFPC2)
wavelengths.}
\label{tab:EROm}
\end{table}

Where are the rest of them?  Either they are not yet assembled, or
they are simply not recognizable.  The latter alternative would be for
example the case in a ``frosting'' (Trager et al.\ 2000) scenario,
where most of the stellar mass is assembled at early-times, while low
levels of star formation contribute the rest of the stellar mass a
later times. In this scenario, some E+S0s would be too blue at
$z\sim1$ to make it into EROs samples (c.f. the range in rest frame UV
colors reported by Moustakas et al.\ 1997 and McCarthy et al.\ 2001
for EROs). Star formation can also alter morphology. For example, if
the secondary star formation activity is concentrated in the disk of
an S0, the disk would become more prominent and active at $z\sim1$,
transforming the S0 into an Sa. This mechanism, of course, would be
effective only on lenticulars and not on pure ellipticals. Therefore
it could perhaps provide an explanation of the deficit of ``old''
spheroids in terms of a demise of lenticulars together with an almost
constant number density of ellipticals (similarly to what is seen in
clusters; Dressler et al.\ 1997; Fasano et al.\ 2000). Deeper,
multicolor, high resolution observations are needed to test whether
the number density evolution of Es and S0s is different.

Let us now turn our attention to semi-analytic hierarchical models.
How do their predictions compare with EROs surface density? Generally,
models significantly under-predict the surface density of EROs. For
example Firth et al.\ (2002) and Smith et al.\ (2002) find 4.5 and 10
times more EROs than expected in the models, a statistically
significant disagreement. Again, we face two possible solutions for
the disagreement. Either hierarchical models do not produce enough
massive systems at $z\sim1$, or simply their colors are wrong. This
latter possibility can occur as a result of an excess of delayed star
formation, or as a result of inappropriate treatment of star formation
and dust extinction in semi-analytic models.

To summarize, it seems that measurements of the basic observable (EROs
surface density; Fig.~\ref{fig:EROs}) is reaching a reasonable level
of mutual agreement. A further improvement would be to gather
redshifts for large number of EROs in order to pin down the effective
redshift selection functions and the three dimensional space density.
Significant efforts are ongoing (Cimatti et al.\ 2002b; Ellis et al.\
2003), and are expected to provide this piece of information soon, at
least at the bright end ($K<19-20$). In spite of these achievements,
the interpretation of the observations is still open. It seems clear
that E+S0s in the local Universe are not all ``old'' (as in a single
burst of star formation at high redshift) and hierarchical models
underpredict the density of EROs. It is still disputed whether this
discrepancy can be accounted for by improving the treatment of star
formation or it is rather pointing to fundamental problems in the {\it
standard model}.

Observations can help addressing this issue in at least two ways. On
the one hand they can provide new constraints less critically
dependent on star formation history (such as the distribution of
redshifts for K-band selected objects described by Kauffmann \&
Charlot 1998 and Cimatti et al.\ 2002a; or the mass focused approach
described in Section~ 3). On the other hand, observations can help by
providing independent and detailed information on the star formation
history of E+S0s.

\section{Star formation history}

We now turn to observational constraints on the star formation history
of E+S0s, particularly those obtained from the redshift evolution of
the color magnitude relation (\ref{sec:cmr}) and the Fundamental Plane
(\ref{sec:FP}). Where possible, I will discuss both cluster and field
observations. Contrasting and connecting the trends across
environments is not only crucial to obtain a complete empirical
picture, but also to test if the formation of E+S0s is delayed in low
density environments as predicted by the {\it standard model}
(Kauffmann 1996).

\subsection{The colors of distant E+S0 galaxies}

\label{sec:cmr}

In the local Universe E+S0s obey a color-magnitude relation (CMR):
brighter E+S0s are redder than less luminous ones. At a given absolute
magnitude, the scatter in optical and infrared colors -- at least in
clusters -- is minimal ($<0.05$ mags; Bower, Lucey \& Ellis 1992). The
widely accepted interpretation is that brighter E+S0s are more metal
rich and that star formation in cluster E+S0s happened and ceased
early enough in cosmic time that the effects of possible spread in
formation epoch are non detectable through broad band colors (see
Bower, Kodama \& Terlevich 1998 for caveats). The most convincing
evidence in support of this interpretation is the redshift evolution
of the CMR. The almost constant slope of the color-magnitude relation
with redshift shows directly that it is not an age-mass sequence
(Ellis et al. 1997; Kodama et al.\ 1998). Similarly, the small scatter
found in high-z clusters indicates that the stellar populations of
massive cluster E+S0s are uniformly old ($z_f>2$) and quiescent
(Stanford, Eisenhardt \& Dickinson 1995, 1998; Ellis et al.\ 1997),
with the possible exception of S0s at large cluster radii (van Dokkum
et al.\ 1998b). What prevents this from being a simple and well
defined picture is that E+S0s in high redshift clusters are not the
only possible progenitors of present day cluster E+S0. Some of the
progenitors at $z\sim1$ might not have yet been accreted onto the
cluster, or might not be morphologically recognizable. Therefore, the
tightness of the CMR in high-z clusters could in part be due to a
selection effect (``progenitor bias''; van Dokkum \& Franx
2001). However, the observed evolution of the morphology-density
relation (Dressler et al.\ 1997; van Dokkum et al. 2001; Treu et al.\
2003) can be used to quantify the bias and rules out the most dramatic
scenarios. 

Less well studied is the CMR in the general field. The few studies
available seem to indicate that there is a CMR in the field out to
$z\sim1$, although with considerably more scatter than in clusters
(Franceschini et al.\ 1998; Kodama, Bower \& Bell 1999; Schade et al.\
1999). Similarly, %\sim 30%\% of the $z>0.5$ early-type galaxies in
the Hubble Deep Fields show strong variations in internal color, often
associated with blue cores (Menanteau, Abraham \& Ellis 2001), at variance with
the homogeneous population and red color gradients (Saglia et al.\
2000) observed in clusters at similar redshifts. 

\subsection{The Fundamental Plane of distant E+S0 galaxies}

\label{sec:FP}

Can we now measure the star formation of history of E+S0s at a given
mass? A promising way to do this is by studying the evolution with
redshift of the Fundamental Plane (Djorgovski \& Davis 1987; Dressler
et al.\ 1987; J{\o}rgensen, Franx \& Kj{\ae}rgaard 1996; hereafter
FP). The FP is a tight empirical correlation between the effective
radius \Rekpc, velocity dispersion $\sigma$, and effective surface
brightness \sbe, of equation:
\begin{equation}
\label{eq:FP} 
\log R_{\rm e} = \alpha \log~\sigma + \beta~SB_{\rm e} + \gamma,
\end{equation} 
where $\alpha$ and $\beta$ are called the slopes, while $\gamma$ is
called the intercept. The very existence of the FP is a remarkable
fact. Any theory of galaxy formation and evolution must be able to
account for its tightness (0.08 rms in $\log R_{\rm e}$). For
discussion of possible physical explanations of the FP relation see
references in Treu et al.\ (2001).

Independent of its origin, the evolution of the FP with redshift can
be linked to the evolution of the stellar mass-to-light ratio of
E+S0s, and hence of their star formation history, in the following
way. Let us define an effective mass $M\equiv \sigma^2 R_{\rm e}$. If
homology holds, i.e. early-type galaxies are structurally similar, the
total mass ${\mathcal M}$ (including dark matter if present) is
proportional to $M$ and the effective mass can be interpreted in terms
of the Virial Theorem (e.g. Bertin, Ciotti \& del Principe 2002). Similarly, an
effective luminosity can be defined as $\log L = -0.4 SB_{\rm e} + 2
\log R_{\rm e} + \log 2\pi$.  Based on these definitions, the $M/L$
(effective mass-to-light ratio) of a galaxy is readily obtained in
terms of the FP observables: $M/L \propto 10^{0.4 SB_{\rm e}} \sigma^2
R_{\rm e}^{-1}$. Using the FP relation to eliminate $SB_{\rm e}$
yields $M/L \propto 10^{-\frac{\gamma}{2.5}}
\sigma^{\frac{10\beta-2\alpha}{5\beta}} R_{\rm
e}^{\frac{2-5\beta}{5\beta}}$

Consider a sample of galaxies at $z>0$ identified by a running index
$i$. The offset of $M/L$ from the local value can be computed as

\begin{equation}
\label{eq:DMLzpar}
\Delta \log(M/L)^i= \Delta \left(\frac{10\beta-2\alpha}{5\beta}\right) \log \sigma^i  + \Delta \left(\frac{2-5\beta}{5\beta}\right)\log {\rm R}_{\rm e}^{i}- \Delta \left(\frac{\gamma^i}{2.5\beta}\right),
\end{equation}
where the symbol $\Delta$ indicates the difference of the quantity at
two redshifts, and $\gamma^i$ is defined as $\log R_{\rm e}^i - \alpha
\log \sigma^i - \beta SB_{\rm e}^i$. For the analysis presented here I
will assume that $\alpha$ and $\beta$ are constant (see Treu et al.\
2001 for discussion). This assumption is consistent with the
observations and makes the interpretation of the results
straightforward. If $\alpha$ and $\beta$ are constant -- and there is
no structural evolution (so that $R_{\rm e}$ and $\sigma$ are
constant) -- then $\Delta \log \left( M/L^i \right)=-\frac{\Delta
\gamma^i}{2.5\beta}$, i.e. the evolution of $\log (M/L^i)$ depends
only on the evolution of $\gamma^i$.  Measuring $\gamma^i$ for a
sample of galaxies at intermediate redshift, and comparing it to the
value of the intercept found in the local Universe, measures the
average evolution of $\log (M/L)$ with cosmic time as $<\Delta \log
(M/L)>=-\frac{< \Delta \gamma>}{ 2.5 \beta }$. If the evolution of the
effective mass-to-light ratio measures the evolution of the stellar
mass-to-light ratio, then the FP becomes a powerful diagnostic of
stellar populations. Not only this diagnostic connects stellar
populations to a dynamical mass measurement, but it is also
intrinsically tight, and thus selection effects are small and can be
corrected (Treu et al.\ 2001; Bernardi et al.\ 2003).

\begin{figure}
    \centering
    \includegraphics[width=10cm,angle=0]{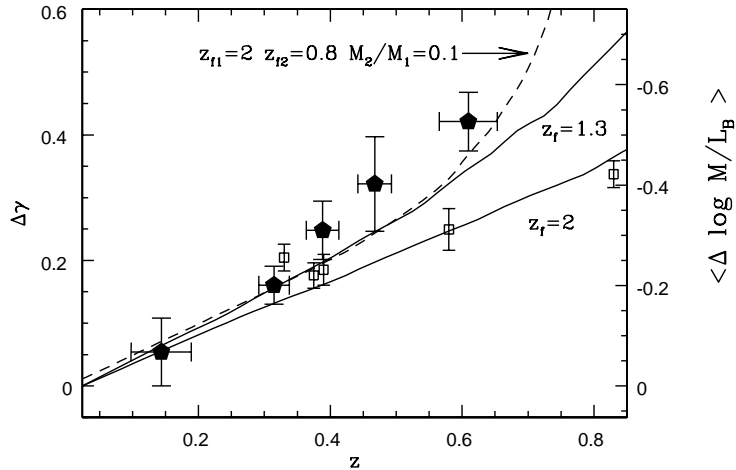}
    \caption{FP in the rest-frame B band. The average offset of the
intercept of field galaxies (Treu et al.\ 2002) from the local FP
relation as a function of redshift (large filled pentagons) is
compared to the offset observed in clusters (open squares; van Dokkum
\& Franx 1996; Kelson et al.\ 1997; Bender et al.\ 1998; van Dokkum et
al.\ 1998a; Kelson et al.\ 2000). The solid lines represent the
evolution predicted for passively evolving stellar populations formed
in a single burst at $z=1.3,2$ (from top to bottom) computed using
Bruzual \& Charlot (1993) models in the BC96 version. The evolution
predicted by a double-burst model is also shown for comparison. See
Treu et al.\ (2002) for details.}
    \label{fig:FP}
\end{figure}

Several groups have applied this technique at $z>0.1$, both in
clusters (Franx 1993; van Dokkum \& Franx 1996; Kelson et al.\ 1997;
Bender et al.\ 1998; van Dokkum et al.\ 1998a; J{\o}rgensen et al.\
1999; Kelson et al.\ 2000; Ziegler et al.\ 2001; van Dokkum \&
Stanford 2003; Fritz et al.\ 2003), and in the field (Treu et
al. 1999; 2001a,b, 2002; van Dokkum et al.\ 2001; Bernardi et al.\
2003). A representative selection of results is shown in
Figure~\ref{fig:FP}. The main results of these studies are: 1) E+S0s
obey a FP relation out to al least $z\sim0.8$ with scatter similar to
local samples (in the field the scatter in $\log R_{\rm}$ is less than
$0.15$ at $z\sim0.8$; Treu et al.\ 2002); 2) field E+S0s (solid
pentagons) evolve faster than cluster ones (open squares). In
quantitative terms, Treu et al.\ (2002) obtain $d \left(\log
M_{*}/L_{\rm B} \right) / dz = -0.72^{+0.11}_{-0.16}$ for the field
sample, while van Dokkum et al.\ (1998a) obtain $-0.49\pm0.05$ for
clusters. Note the good agreement with the results obtained by I02
from the evolution of the LF (\ref{sec:LF}).

In terms of evolution of stellar populations, the cluster data are
consistent with passive evolution of an old stellar population
($z_f\sim2$). A more recent ``epoch of formation'' $z_f\sim1.3$
appears to be needed to explain the field E+S0s evolution in terms of
single burst stellar populations. However, as for the EROs PLE models,
it is sufficient to rejuvenate an old stellar population with a small
amount of recent star formation to obtain an evolution consistent with
the data. For example, the data are well described by a model where
90\% of the stellar mass is formed at $z_{f1}\sim2$ and secondary
bursts at $z_{f2}\sim1$ contribute the residual 10 \%.

\subsection{Discussion}

Both the evolution of colors and FP to $z\sim1$ appear to be
consistent with the following picture. Massive cluster E+S0s are old
and quiescent, while field examples show some relatively recent star
formation activity. This picture is also in qualitative agreement with
the fact that {\it at any given morphological type} star formation
activity decreases monothonically with (local) galaxy density (see,
e.g. Poggianti et al.\ 1999; Poggianti, Dressler \& Nichol, these
proceedings), and observations of high redshift E+S0s based on other
spectroscopic diagnostic features (Schade et al.\ 1999; Kelson et al.\
2001; Treu et al.\ 2002). Further support for this picture comes
from the fossil evidence (Bernardi et al.\ 1998; Trager et al.\ 2000;
Kuntschner et al.\ 2002) , although interpreting the observations in
the local Universe is more difficult, because possible differences
could have been quenched by time to a level where uncertainties on
dust extinction and absolute distances (Pahre, Djorgovski \& de
Carvalho 1998), and the age/metallicity degeneracy (Kuntschner et al.\
2002) are significant.

How does this picture compare with CDM predictions? {\it
Qualitatively}, the observational picture is similar to theoretical
predictions (Diaferio et al.\ 2001; Benson et al.\ 2002). However,
{\it quantitatively}, the observed differences between the star
formation history of field and cluster E+S0s are smaller than predicted
by models. Whereas observations indicate at most minor departures from
a single old stellar populations, hierarchical models predict dramatic
differences already at $z<0.5$ (see Kauffmann 1996 and van Dokkum et
al.\ 2001). As it was the case for EROs, improvements in the treatment
of star formation or of environmental effects might reconcile the
model with the data. Alternatively, this might prove a major problem
for the {\it standard model}, especially when more precise
measurements will be available. From on observational point of view,
it has to be noticed that current studies are based on few tens of
objects at most. It is now crucial to collect high quality data on
larger numbers of distant E+S0s to overcome small sample statistics and
cosmic variance.

\section{The mass density profile of distant E+S0 galaxies}

So far, in this review, I have interpreted observations in terms of
pure luminosity evolution. For example, when expressing the evolution
of the FP in terms of evolution of stellar mass-to-light ratio, I have
assumed pure luminosity evolution. Is there any way we can relax this
assumption and measure directly and simultaneously the internal
structure and stellar populations properties of distant E+S0s? If we
could, not only we could test if the results obtained under a pure
luminosity evolution hypothesis are correct, but also, and most
importantly, we could gain new and fundamentally different insight
into the evolution of E+S0s. For example, measuring the mass density
profile of luminous and dark matter in E+S0s as a function of redshift
not only yields an independent determination of the evolution of the
stellar mass to light ratio, but also tests the existence of the
universal dark matter profile predicted by the {\it standard model}.
Furthermore, theoretical predictions related to the mass density
profile and orbital structure might not be as dramatically sensitive
to the details of the treatment of star formation as, e.g., EROs
number density. Therefore, testing these predictions might be a more
robust way to test the {\it standard model}.

Measuring the mass density profile of E+S0s is already challenging in
the local Universe (e.g. Bertin et al.\ 1994), and traditional methods
are inapplicable at high redshift (for example surface brightness
dimming prevents the measurement of very extended velocity dispersion
profiles). Nevertheless, mass density profile measurements at high
redshifts are possible because distant E+S0s are efficient
gravitational lenses. The next two sections describe recent results on
the mass density profile of distant E+S0s from weak lensing
(\ref{sec:weak}), and joint strong lensing and dynamical analysis
(\ref{sec:strong}).

\subsection{Galaxy-galaxy lensing}

\label{sec:weak}

The distortion of background galaxies lensed by an individual E+S0s is
not detectable. However, if several (at least hundreds) E+S0s are
considered, and the signal from all the background objects is coadded,
a statistical measurement of mass density profile of the average
galaxy can be derived (Brainerd, Blandford \& Smail 1996). 
This technique is known
as galaxy-galaxy lensing and has proved viable to study the outer
regions of the dark matter halos of E+S0s (Griffiths et al.\ 1996).
For example, dark matter halos around red galaxies have been detected
out to several hundreds kpc in SDSS images (McKay et al.\
2003). Combining information from galaxy-galaxy lensing with the
existence of the FP, Seljak (2002) showed that at large radii the mass
density profile of E+S0s declines faster than $r^{-2}$, consistent
with the $r^{-3}$ behavior predicted by CDM numerical
simulations. Natarajan \& Kneib (1997) and Natarajan et al.\ (1998)
showed that dark matter halos of E+S0s can be detected even within
clusters if the cluster potential is appropriately modeled. Natarajan,
Kneib \& Smail
 (2002) applied this technique to WFPC2 images of a sample of
intermediate redshift clusters, and showed that dark matter halos of
E+S0s are truncated as expected from tidal interaction with the
cluster.
 
\subsection{Strong lensing and the Lenses Structure and Dynamics Survey}

\label{sec:strong}

The majority of the almost hundred galaxian gravitational lenses known
are E+S0s. Once the redshift of the lens and the source are known, the
geometry of the multiple images provides a very robust measurement of
the mass enclosed by the Einstein Radius $R_{\rm E}$. The Einstein
Radius of the typical $z\sim0.5$ E+S0s lens galaxy is larger than the
effective radius. Thus strong lensing can be used to determine total
mass at large radii for tens of distant E+S0s, independent of the
nature and dynamical state of the mass inside $R_{\rm R}$.

\begin{figure}
    \centering \includegraphics[width=4.5cm,angle=0]{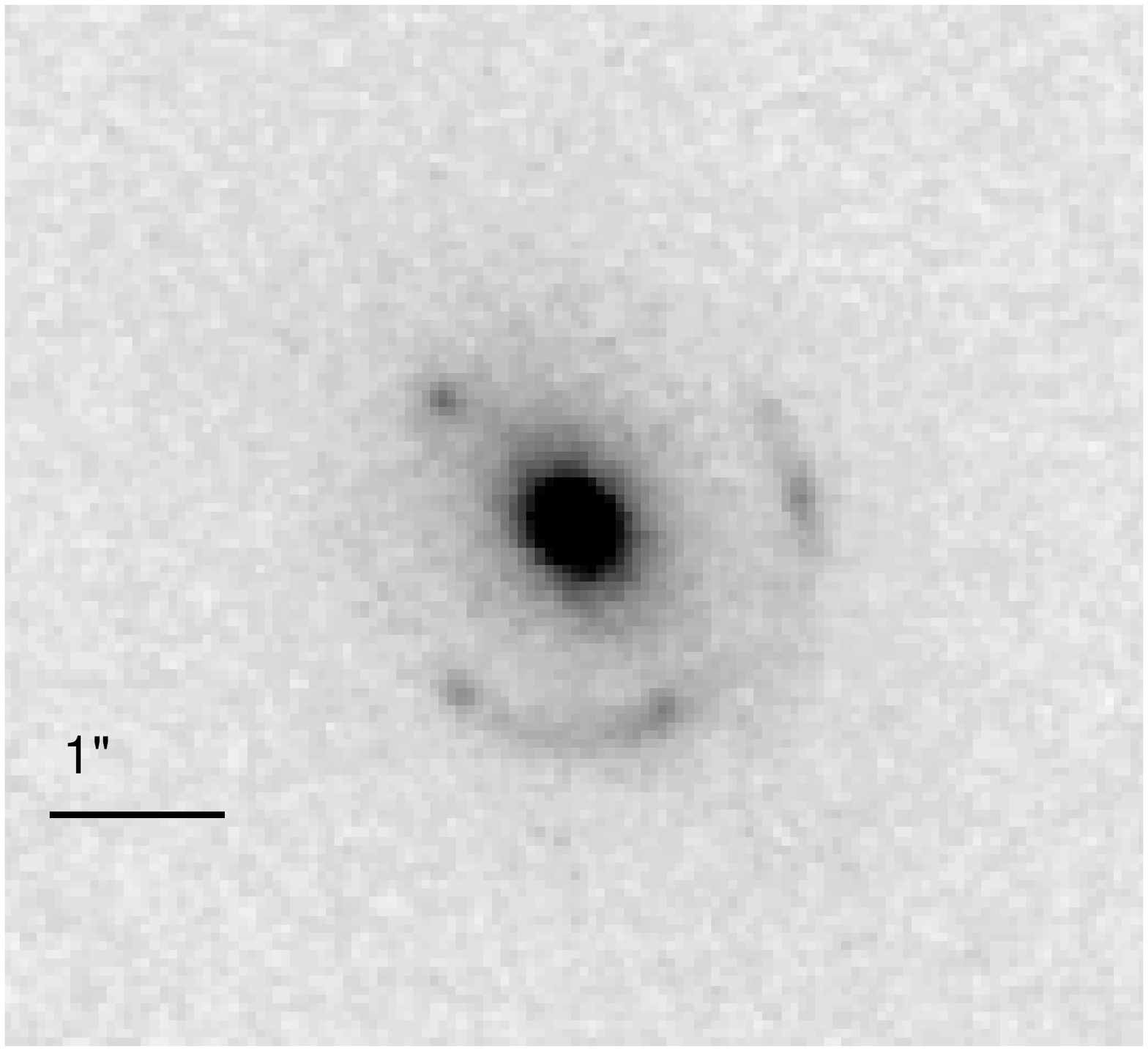}
    \includegraphics[width=6.5cm,angle=0]{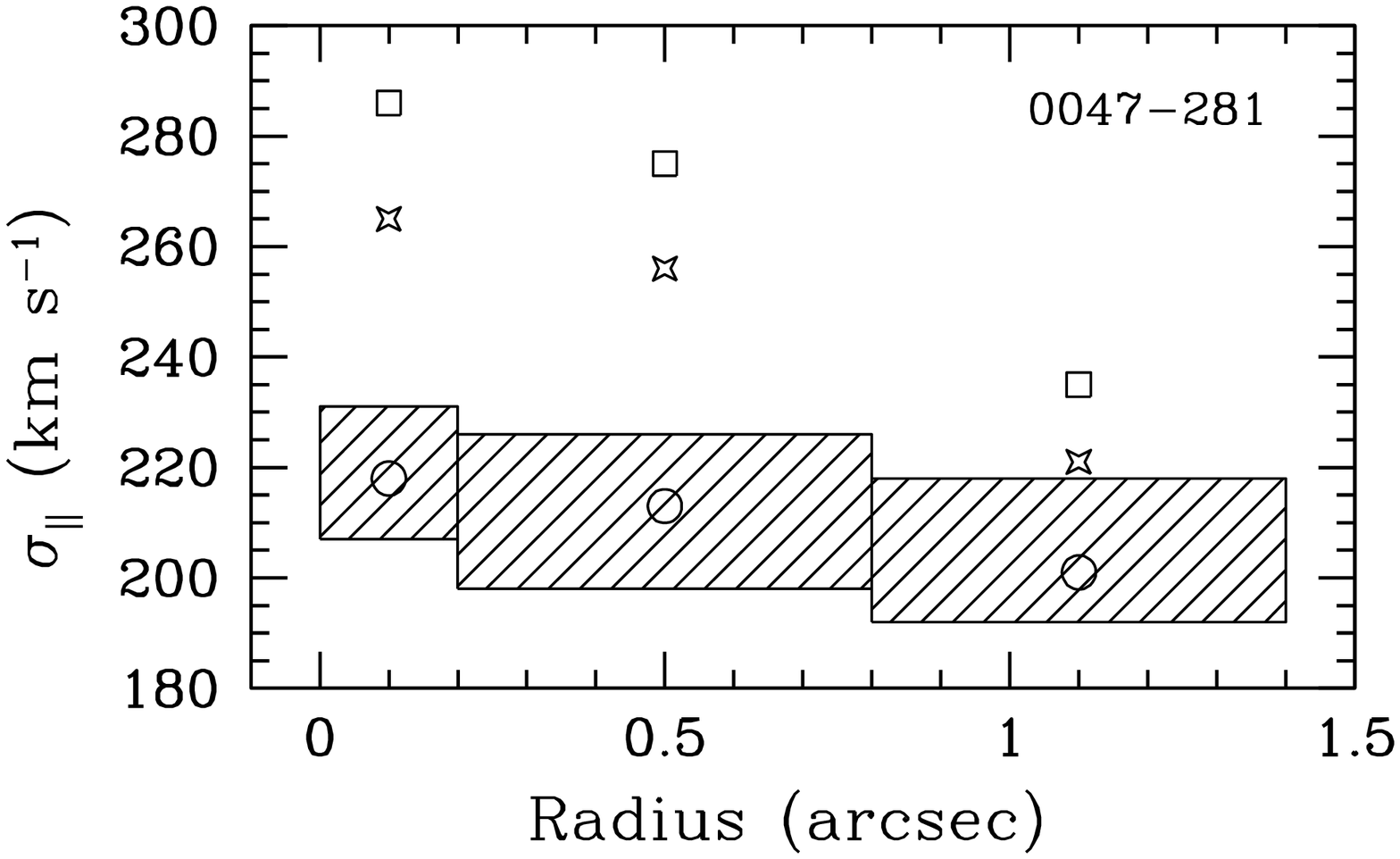}
\caption{Left: HST image of 0047$-$281 at $z=0.485$. Right: velocity
dispersion profile of 0047$-$281 along the major axis. The box height
indicates the 68\% measurement error, whereas the box width indicates
the spectroscopic aperture. The open squares are the corresponding
values for an isotropic constant M/L model, which is rejected by the
data. See Koopmans \& Treu (2003) for details.
\label{fig:H0047}}
\end{figure}

In some cases, knowledge of the mass enclosed by $R_{\rm E}$ is
already sufficient to show that the average total mass-to-light ratio
is larger than expected for reasonable stellar populations, and
therefore to prove the existence of dark matter.  Unfortunately, not
much information is generally provided on how mass is spatially
distributed\footnote[4]{Except perhaps when the lensed source is
extended and the detailed geometry can be used to increase the number
of constraints (see, e.~g., Blandford et al.\ 2001; Kochanek et al.\
2001; Saha \& Williams 2001)}.

Nevertheless, assuming a mass density profile, lensing can be used to
probe the evolution of the stellar populations. For example, Kochanek
et al.\ (2000) and Rusin et al.\ (2003) used image separation to
estimate the velocity dispersion of lens E+S0s assuming a singular
isothermal total mass density profile (i.e. the total density
$\rho_t\propto r^{-2}$). With this assumption they measure the
evolution of the FP of lens galaxies and find $d \left(\log
M_{*}/L_{\rm B} \right) / dz = -0.54 \pm 0.09$, i.e. more similar to
the cluster value than the field value (a similar analysis of lens
E+S0s by van de Ven et al.\ (2003), yields $-0.62\pm0.13$). The
marginally significant differences with respect to the direct method
could be the result of different selection processes (lenses are
``mass'' selected, while samples used in direct measurements are
``light'' selected), of different environments (lenses might be
preferentially found in groups or small clusters; Fassnacht \& Lubin
2002), or of external contributions to the image separation (such as
from a nearby group or cluster). Or perhaps, the differences could be
an indication of small departures from isothermal mass density
profiles. However, the ability of a simple singular isothermal mass
model to predict with reasonable accuracy the central velocity
dispersion is remarkable (as generally confirmed by direct
measurement, e.g., Koopmans \& Treu 2002). This is an indication of
the overall structural homogeneity of E+S0s. The accuracy of the
predicted $\sigma$ is even more remarkable considering that $R_{\rm
E}/R_{\rm e}\sim0.5-5$, and therefore lensing probes regions dominated
by stellar mass as well as regions dominated by dark matter.

More can be learned on the internal mass distribution of distant E+S0s
by combining strong lensing constraints with spatially resolved
stellar kinematics of the lens galaxy, in a joint lensing and
dynamical analysis. The two diagnostics complement each other reducing
the degeneracies inherent to each method.  Stellar kinematics
constrains the mass distribution within the Einstein radius, while
gravitational lensing analysis fixes the mass at the Einstein radius,
thus lifting the so-called mass-anisotropy degeneracy (Treu \&
Koopmans 2002, Koopmans \& Treu 2003).

Combining the two diagnostics is the goal of the Lenses Structure and
Dynamics Survey (Koopmans \& Treu 2002,2003; Treu \& Koopmans 2002a,
b; hereafter collectively KT). In eight clear nights at the Keck-II
Telescope we have collected data to measure accurate and spatially
resolved stellar kinematics for a sample of 11 gravitational lenses
out to $z\sim1$ with available HST images. An example of the data is
shown in Figure~\ref{fig:H0047}.

A family of two-component spherical mass models is used in the joint
lensing and dynamical analysis (see KT for details). One component is
the stellar component, assumed to follow the surface brightness
profile as measured from HST images scaled by a constant stellar mass
to light ratio ($M_*/L_B$). The other component is the dark matter
halo, modeled as a generalized NFW profile, where the dark matter
density goes as $r^{-3}$ at large radii and $r^{-\gamma}$ at small
radii. Comparison with the data yields best fitting models and
likelihood contours on the relevant parameters ($M_*/L_B$ and
$\gamma$).

\begin{figure}
    \centering \includegraphics[width=6.0cm,angle=0]{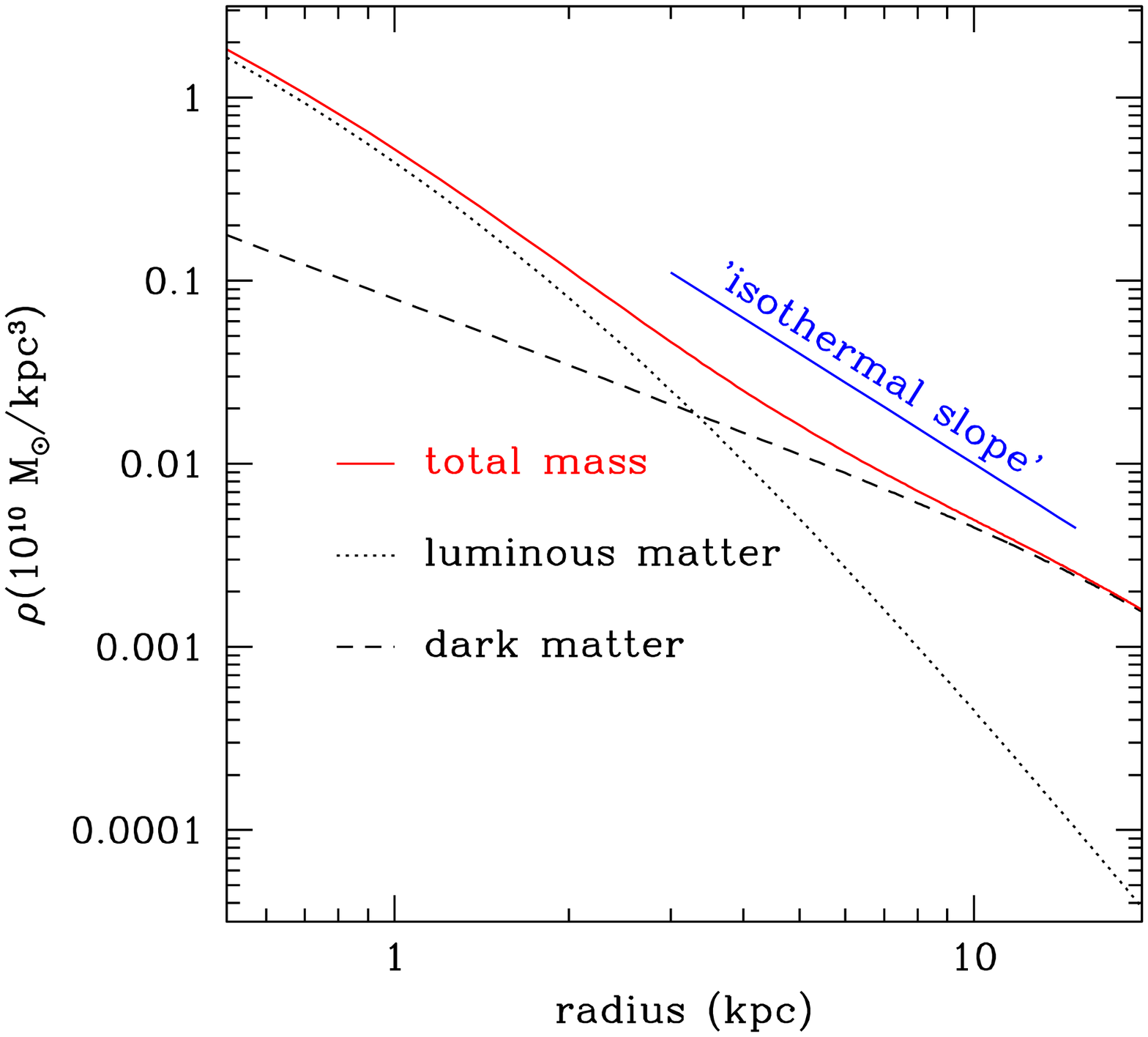}
    \includegraphics[width=5.5cm,angle=0]{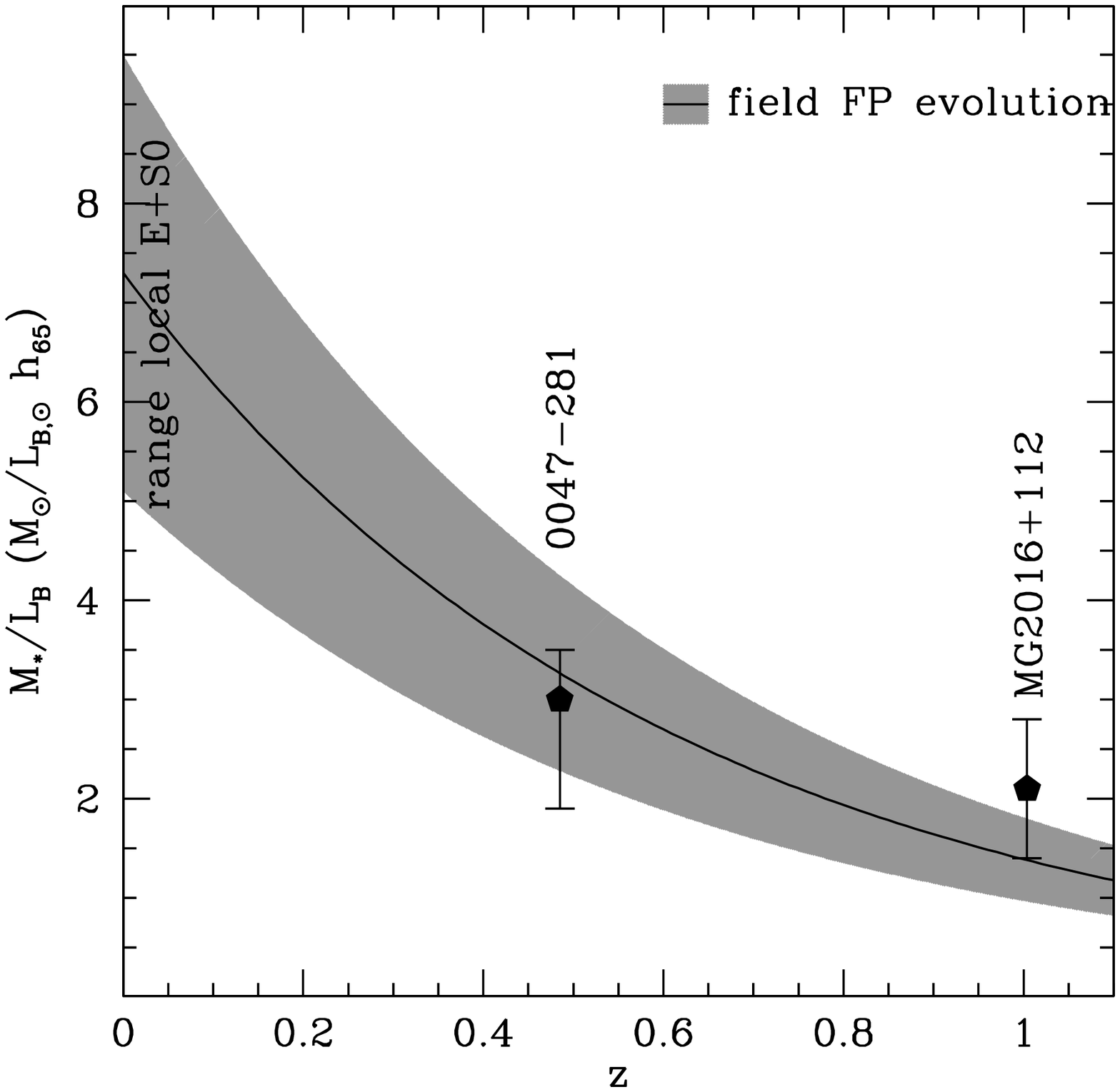} \caption{Left:
    best fitting mass model of MG2016+112 at $z=1$ (see Treu \&
    Koopmans 2002a for details). Right: comparison between the
    evolution of stellar the mass to light ratio measured via the FP
    evolution and via a joint lensing and dynamical analysis by the
    LSD Survey (see text for detail).}  \label{fig:MG2016}
\end{figure}

The best fitting mass model for MG2016+112 at $z=1.004$ (Lawrence et
al.\ 1984) is shown in the left panel of
Fig.~\ref{fig:MG2016}. Luminous mass dominates in the inner $\sim10$
kpc while a flatter dark matter halo contributes most of the mass at
larger radii. No dark matter models, or constant total mass-to-light
ratio models, are rejected at high confidence level.  Remarkably,
although none of the two components is a simple power law density
profile, the {\it total} mass density profile follows very closely an
$r^{-2}$ singular ``isothermal'' profile (equivalent to a flat
rotation curve for spiral galaxies). The same result is recovered by
modeling the lens with a simple power law mass density profile
$\rho_t\propto r^{-\gamma'}$. Comparison with the data yields
$\gamma=2.0\pm0.1\pm0.1$. Similar results are found for the other
object analyzed so far, 0047-281 (Warren et al.\ 1996) at
$z=0.485$. There is strong evidence for a dark matter halo more
diffused than the luminous component and the total mass density
profile is close to a singular isothermal profile, well described by a
power law with effective slope $\gamma'=1.90^{+0.05}_{-0.23}\pm0.1$.

The joint lensing and dynamical analysis also yields a measurement of
$M_*/L_B$, which can be used to measure the evolution of stellar
population independently of the FP analysis (\ref{sec:FP}). In the
right panel of Fig.~\ref{fig:MG2016} the LSD results are compared with
the FP results. Since the FP analysis only yields \dlogMLB, I adopted
for the comparison the $7.8\pm2.7$ in solar units for the range of
local values (KT). The agreement is very good, consistent with the
expectations of a pure luminosity evolution scenario.

Finally, observational limits on $\gamma$ can be used to test the
cuspy dark matter halos predicted by CDM scenarios ($\gamma=1$ NFW;
$\gamma=1.5$ Moore et al.\ 1998). For the first two objects we find
upper limits $\gamma<1.4$ and $\gamma<1.5$ (68\% CL) consistent with
the results of numerical simulations only if the collapse of baryons
to form stars (which dominates in the inner regions) did not steepen
significantly the dark matter halos. The analysis of the complete
sample will hopefully provide more stringent limits.

Although the results so far have to be considered preliminary since
they are based on the first two objects, three facts appear to stand
out: i) E+S0s at high redshift have diffuse dark matter halos; ii)
luminous and dark matter, although spatially segregated, ``conspire''
to follow an almost isothermal total mass density profile, similarly
to what happens in local E+S0 and spiral galaxies (van Albada \&
Sancisi 1986; Rix et al.\ 1997); iii) the agreement between the
evolution of the mass to light ratio measured by the FP and the direct
measurements is consistent with no structural evolution of E+S0s in the
past 8 Gyrs.

The first point, direct evidence of extended dark matter halos around
E+S0s out to $z\sim1$, is probably not surprising, but it is a
confirmation of the CDM scenario. The second point, the ``conspiracy''
between luminous and dark matter to produce $r^{-2}$ -- that appears
to be a consistent feature of early-type and spiral galaxies out to
$z\sim1$ -- is something that should be explained by a satisfactory
cosmological model. It is not clear if this is the case in the {\it
standard model}, since simulations do not generally include baryons.
Analytic approximations of baryonic collapse (Blumenthal et al.\ 1986)
do not explain naturally this result. An alternate explanation might
be that the $r^{-2}$ profile, the limit of (incomplete) violent
relaxation (Lynden-Bell 1967), is a {\it dynamical attractor}. If
baryons are transformed in stars early enough (as in the monolithic
collapse scenario by van Albada 1982 or as recently proposed by Loeb
\& Peebles 2003), then they behave as dissipationless particles and
could interact with dark matter so as to tend to a {\it total} mass
density profile that is close to the dynamical attractor (Loeb \&
Peebles 2003), while preserving spatial segregation as a result of
different initial conditions.  Finally, the third point, the lack of
dynamical evolution out to $z\sim1$, together with other evidence for
homogeneity of E+S0s described in the previous sections, appears to be
another challenge for the {\it standard model}. If E+S0s are formed by
mergers, then either mergers have to occur very early in cosmic time,
or some sort of fine tuning of the merging process appears to be
required in order to produce such homogeneous end-products.

\section{Acknowledgments}%

I am grateful to Giuseppe Bertin and Richard Ellis for their comments
on an earlier version of this manuscript, and to my collaborators
Stefano Casertano, L\'eon Koopmans, Palle M{\o}ller, and Massimo
Stiavelli for innumerous stimulating discussions. I acknowledge useful
conversations with Masataka Fukugita, Myungshin Im, Pat McCarthy,
Alvio Renzini, David Sand, Graham Smith. I would like to thank the
organizers for this exciting meeting, and the referee, Alan Dressler,
for insightful comments.

\end{document}